\documentclass[]{iopart}
\pdfoutput=1
\usepackage{lmodern}
\usepackage{iopams,iopsetstack}
\usepackage{fixltx2e} 
\usepackage[T1]{fontenc}
\usepackage[utf8]{inputenc}
\IfFileExists{upquote.sty}{\usepackage{upquote}}{}
\IfFileExists{microtype.sty}{%
\usepackage{microtype}
\UseMicrotypeSet[protrusion]{basicmath} 
}{}
\usepackage[unicode=true]{hyperref}
\usepackage[usenames,dvipsnames]{color}
\hypersetup{breaklinks=true,
            bookmarks=true,
            pdfauthor={},
            pdftitle={Ordering and Dynamics of Vibrated Hard Squares},
            colorlinks=true,
            citecolor=blue,
            urlcolor=blue,
            linkcolor=magenta,
            pdfborder={0 0 0}}
\usepackage{graphicx,grffile}
\makeatletter
\def\maxwidth{\ifdim\Gin@nat@width>\linewidth\linewidth\else\Gin@nat@width\fi}
\def\maxheight{\ifdim\Gin@nat@height>\textheight\textheight\else\Gin@nat@height\fi}
\makeatother
\setkeys{Gin}{width=\maxwidth,height=\maxheight,keepaspectratio}
\usepackage{booktabs}
\newcommand{\abs}[1]{\left| #1 \right|}
\newcommand{\ang}[1]{\left\langle #1 \right\rangle}

\begin{document}
\title{Ordering and Dynamics of Vibrated Hard Squares}
\author{Lee Walsh\(^{1}\) and Narayanan Menon\(^{1,2}\)}
\address{\(^1\)Department of Physics, University of Massachusetts, Amherst, USA}
\address{\(^2\)TIFR Centre for Interdisciplinary Science, Hyderabad, India}
\ead{\mailto{lawalsh@physics.umass.edu}, \mailto{menon@physics.umass.edu}}

\begin{abstract}
We study an experimental system of hard granular squares in two
dimensions, energized by vibration. The interplay of order in the
orientations and positions of anisotropic particles allows for a rich
set of phases. We measure the structure and dynamics of steady states as
a function of particle density. This allows us to identify a progression
of phases in which a low density isotropic fluid gives way to a phase
with tetratic orientational order, short-range translational
correlations, and slowed rotational dynamics. In this range of density
we also observe a coupling between the molecular orientational order and
bond-orientational order. At higher densities, the particles freeze into
a translationally and orientationally ordered square crystalline phase
in which translational diffusion is suppressed.
\end{abstract}
\begin{indented}
\item[]{\bfseries Keywords:}\enskip\ignorespaces
{\footnotesize granular matter, phase diagrams (experiments), jamming and packing, driven diffusive systems (experiments)}
\end{indented}

\section{Introduction}\label{introduction}

An assembly of hard squares in two dimensions presents a variety of
potential spatial orders. Hard particles form thermodynamic phases
dictated solely by the entropy of their geometric packing; temperature
is irrelevant as the interaction potential has no finite energy scale.
Particles with asymmetric shapes have anisotropic constraints of
self-avoidance and therefore pack in complex phases. Squares are an
obvious point of entry into this complexity because they can tile the
plane without frustration. However, there has been relatively little
investigation into the phases formed by this simple shape as a function
of density, and still less of the dynamics within these phases.

The phase diagram of isotropic hard disks~{[}1--3{]} offers a context
for the ordering of asymmetric particles. A two-dimensional crystal of
disks has correlations in particle positions with algebraic (power-law)
decay. As density is decreased the crystal melts into the hexatic phase
in which the orientation of ``bonds'' between neighboring particles has
algebraic correlations, which finally melts into an isotropic liquid
phase. Anisotropic particles have both translational and rotational
degrees of freedom, each of which may order independently or become
coupled, providing the possibility of new classes of order. For example,
particle (or ``molecular'') orientations may order independently of
their locations, giving rise to phases with global orientational order
in the absence of translational order, e.g.~the nematic liquid crystal
phase. These phases in two dimensions have been explored using various
rigid shapes with a two-fold rotational symmetry, such as
rectangles~{[}4,5{]}, ellipses~{[}6{]}, or rods~{[}7--9{]}. Some of
these two-fold symmetric particles yield phases with four-fold symmetry.
This has also been demonstrated by density functional
theories~{[}10,11{]}. Minor differences in shape can have drastic
effects on the phase diagram of such systems in two
dimensions~{[}12--14{]}, and three dimensions~{[}15--17{]}.

Existing work on shapes with four-fold symmetry --- such as squares ---
is limited, but raises interesting questions about the phase diagram.
Monte Carlo simulations of an equilibrium hard-square system by
Wojciechowski and Frenkel suggest a four-fold tetratic phase with
longer-range orientational order than translational order~{[}18{]}. By
contrast, in experiments on a colloidal suspension of square-shaped
tiles osmotically constrained to a surface, Zhao~\emph{et~al.} found no
evidence of an orientationally ordered state~{[}19{]}. At densities
above that of the isotropic phase, they instead observed a ``hexagonal
rotator'' phase where particles form a plastic crystal with hexagonal
bond-orientational order but no molecular-orientational order. At higher
densities, the system enters a rhombic phase with a lattice angle that
continuously approaches the 90° angle of a square crystal at full
packing.

\begin{table}[h]
\centering
\begin{tabular}{@{}cccl@{}}
\toprule
Molec. & Bond  & Transl. & Phase  \\ \midrule
---    & ---   & ---     & isotropic liquid                     \\
X      & ---   & ---     & 4-fold molecular orientational order \\
---    & X     & ---     & 4-fold bond-orientational order      \\
X      & X     & ---     & tetratic                             \\
---    & X     & X       & plastic (rotator) crystal            \\
X      & X     & X       & square crystal                       \\ \bottomrule
\end{tabular}
\caption{Potential phases from combinations of translational, bond-, and
molecular-orientational order. ``---'' indicates short-range order with
exponentially decaying spatial correlations; ``X'' indicates quasi-long-range
order with algebraic decay.}
\label{tab:intro}
\end{table}

Monte Carlo simulations by Avendaño and Escobedo~{[}12{]} indicate that
particle shape is the critical difference between these two studies.
They found that squares with sufficiently rounded corners form the
hexagonal rotator phase observed by Zhao,~\emph{et~al.}~{[}19{]},
whereas squares with sharper corners show indications of a tetratic-like
phase before crystallization~{[}18{]}.

Our present study is motivated by the absence of experiments in the hard
square limit. Furthermore, individual particle dynamics can be measured
via experiment, which Monte Carlo simulations do not capture. In this
work we present experimental observations of structure and translational
and rotational dynamics of the steady states of a system of vibrated
hard granular squares. The macroscopic nature of the system gives us the
advantage of precise control over the particle properties, including
shape~{[}8,20--22{]}. This allows us to achieve the limit of
sufficiently sharp corners to accurately represent hard
squares~{[}12{]}. Our system size is necessarily small, though
comparable to previous simulations and experiments~{[}9,12,18,19{]}. As
a result, transitions are rounded off due to finite number, and
long-range correlations are truncated by the boundary.

We find a progression from an isotropic disordered fluid at low density,
through a fluid with orientational order, arriving at a solid with both
orientational and translational order. The intermediate orientational
order is tetratic, in which bond- and molecular-orientational order are
coupled and each has four-fold symmetry. Neither hexagonal rotator nor
rhombic phases are seen. The static structure analysis gives qualitative
agreement with Monte Carlo simulations~{[}12,18{]}. We also present new
results on the dynamics of particles as a function of density. We find
that rotational diffusion is suppressed relative to translational
diffusion in the approach to the tetratic phase. At still higher
densities, the transition to the solid is accompanied by the loss of
translational diffusion.

\begin{figure} \center
\includegraphics[width=.8\textwidth]{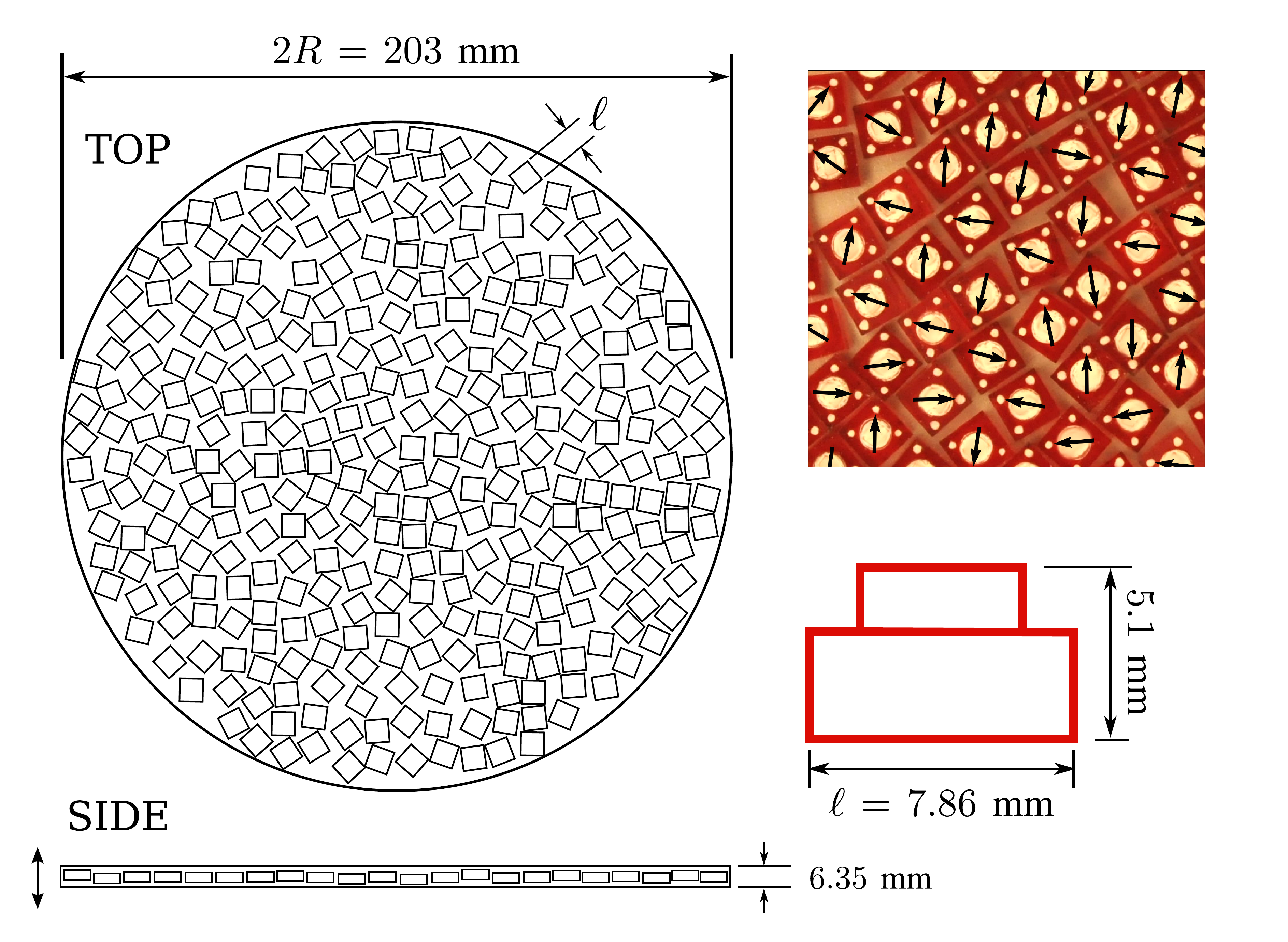}
\caption{Experimental setup. Particles ($N \approx 500$) are constrained in a quasi-two-dimensional circular dish between an aluminum substrate and acrylic cover. The dish has diameter $2R = 203$ mm with height 6.35 mm, and particles have side length $\ell = 7.86 \pm 0.01$ mm and height 5.1 mm. Photo at upper right shows the particles (at density $\rho=0.78$) painted with white tracking targets at center and three corners.
Arrows show location and orientation of detected particles.}
\label{fig:expt}
\end{figure}

\section{Experiment}\label{experiment}

We confined square particles to a monolayer in a horizontal dish.
Vertical vibration provides quasi-thermal noise in two dimensions and
the particles diffuse in plane. The setup is shown in
figure~\ref{fig:expt}. Particles are hard plastic tiles (manufactured by
LEGO) with sides of length \(\ell = 7.86 \pm 0.01\)~mm and base height
of 3.2~mm, plus a 1.9~mm high cylindrical protrusion on the top, for a
total height of 5.1~mm. Corners have radius of curvature
\(\sigma/2 = 0.15 \pm 0.01\)~mm, giving a corner-rounding-to-length
ratio \(\zeta = \sigma / \ell = 0.04\)~{[}12{]}. Their mass is
\(204 \pm 1\)~mg. The particles are contained in a circular aluminum
dish of diameter \(2R = 20.3\)~cm with an acrylic lid that maintains a
vertical gap of 6.35~mm (1.25 particle height). The dish is coupled to a
permanent magnet electrodynamic shaker (LDS~V456~shaker and
PA1000L~amplifier), which vibrates the dish vertically following
sine-wave motion with peak acceleration \(\Gamma \approx 10 g\) at
frequency \(f = 50\)~Hz.

Particles were imaged from above with a high-speed video camera (Vision
Research Phantom~v7.1) to measure displacements, and a DSLR camera
(Nikon~D5000) to measure positions. They are marked with dots for
detection of position and orientation, as shown in
figure~\ref{fig:expt}. For dynamic measurements, we uniquely identified
and tracked particles through \(120\)-fps video. Positional uncertainty,
primarily from lighting and imaging, is \(0.25\) pixels, which
corresponds to \(0.01 \ell\) in video images and \(0.002 \ell\) in still
images, where \(\ell\) is the particle size. For all data, orientational
uncertainty is \(0.02\) rad and is dominated by inaccuracy in marking
dot location.

We controlled the density (packing fraction) \(\rho\) by varying the
number of particles \(N\) in the dish. Particles tend to align with the
dish walls, therefore a margin of two particle layers is excluded from
the analysis and density calculations. The choice of the margin size
affects the density by up to \(\pm 0.004\), while the temporal
fluctuations of the density are of order \(\pm 0.005\).

In this nonequilibrium system, energy is fed to particles via collisions
with the vibrating cell. Collisions between particles and the bottom
plate, the lid, and other particles are inelastic, and kinetic energy is
dissipated quickly. Likewise, momentum of the particles is not conserved
due to collisions and friction with the substrate. The nonequilibrium
drive from the vertical vibration provides noise that leads to motion in
the horizontal direction. The symmetry of the resultant impulse on the
particle matches the symmetry of the particle shape, but the long-term
motion of the tiles is isotropic.

\section{Structure}\label{structure}

In this section, we study the emergence of ordered phases by capturing
static images of the system over a range of densities. The images were
taken at time intervals (minutes) long enough compared to typical
diffusion times (seconds) to obtain statistically independent
configurations of the particles. We characterized spatial order with
order parameters and correlation functions computed from particle
positions and orientations.

\subsection{Positional Order}\label{positional-order}

\begin{figure} \center
\includegraphics[width=.8\textwidth]{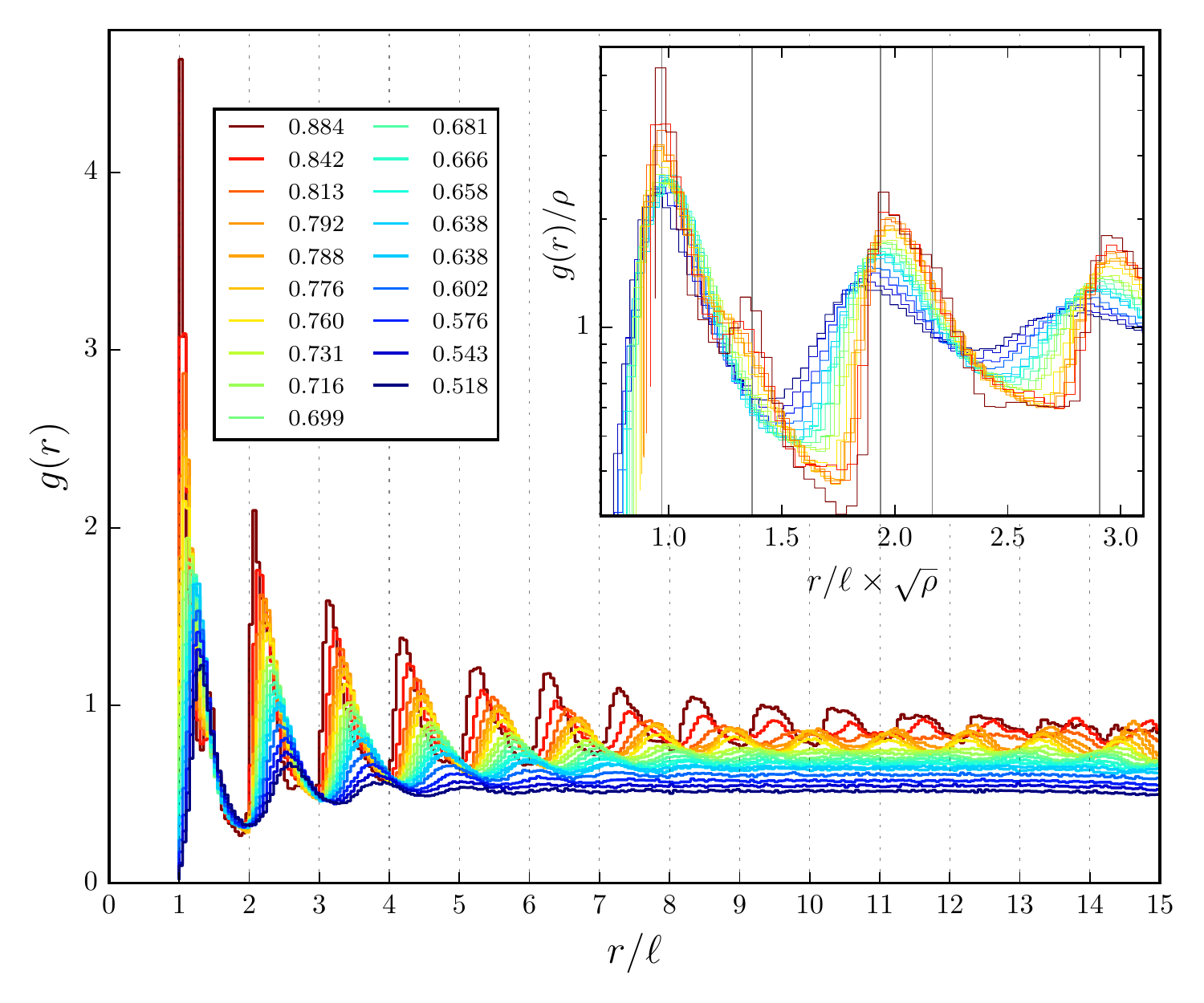}
\caption{The radial distribution function $g(r)$ at various densities. The spatial range of translational order increases with density. Inset: Detail of the nearest peaks, rescaled by density to show the evolution of square structure as the density is increased. Vertical lines mark $1, \sqrt 2, 2, \sqrt 5,$ and 3 times the average inter-peak spacing. The developing peaks and shoulders near $\sqrt 2$ and $\sqrt 5$ indicate the emergence of square-crystalline order around densities $\rho \gtrsim 0.78$.}
\label{fig:gr}
\end{figure}

\begin{figure} \center
\includegraphics[width=\textwidth]{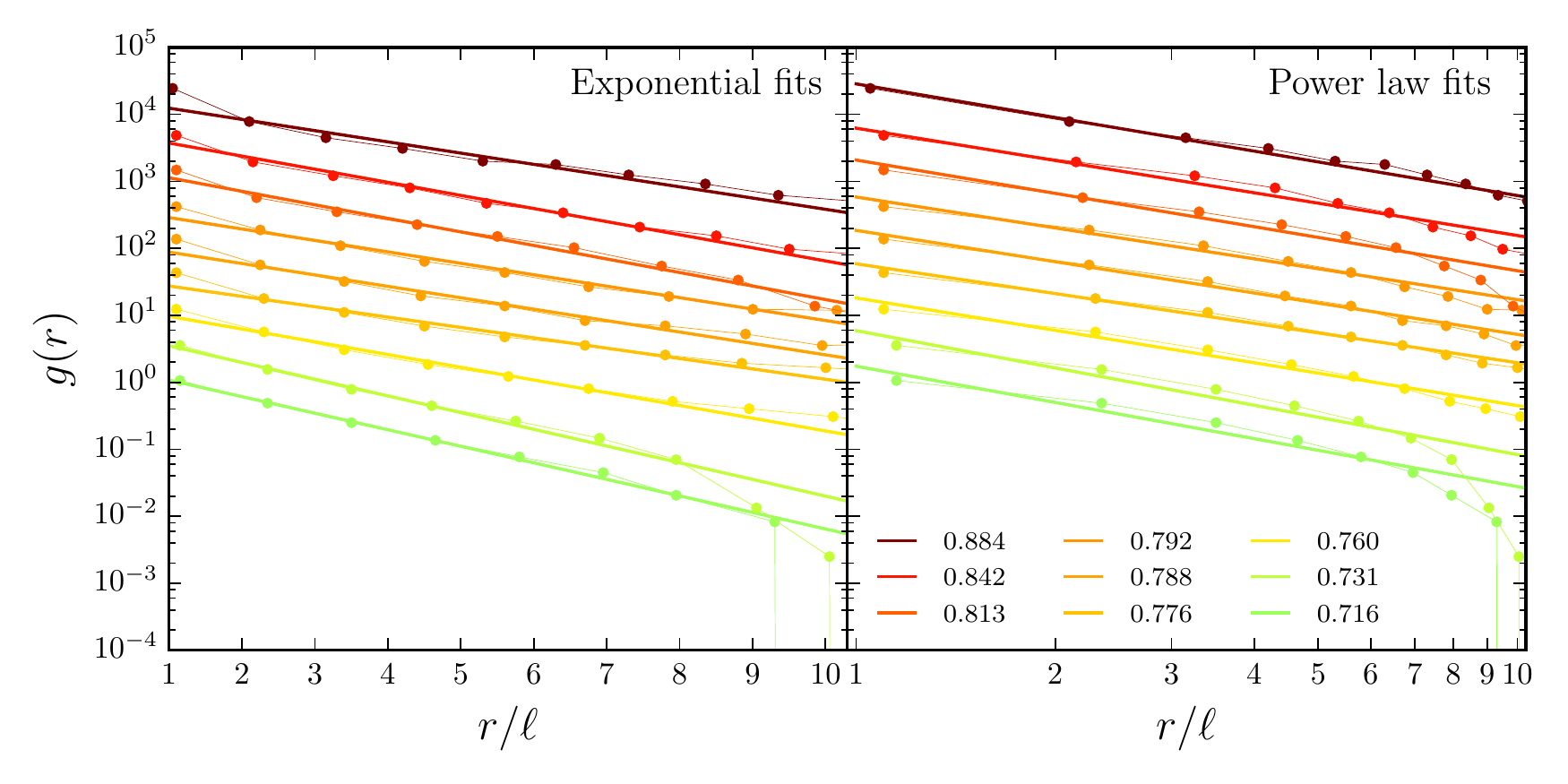}
\caption{Fits to g(r) peaks. The decay shape of the peak envelope for $g(r)$ shows the extent of translational order. Shown are the peaks from $g(r)$ with fits to exponential\ (left) and power-law (right) decay functions. Curves of different density are vertically offset by a factor of 3 for clarity; the lowest curve has no offset.}
\label{fig:grfits}
\end{figure}

We first consider translational order by studying particle positions,
regardless of their orientations. Translational order can be quantified
by the radial distribution function~\(g(r)\), which gives the
probability density of finding two particles separated by
distance~\(r\). Our measurement is shown in figure~\ref{fig:gr}. As
density increases, we observe sharper and taller peaks; the peaks are
asymmetric, with sharp lower bounds at integer values of~\(r/\ell\),
which correspond to particle separation in a perfect square packing. At
high densities (\(\rho \gtrsim 0.78\)), a shoulder and then a new peak
begins to form near \(r/\ell = \sqrt 2\) corresponding to the
second-nearest neighbor distance in a square crystal.

The characteristic length of translational order can be determined from
the functional form and length scale of the decay envelope of~\(g(r)\).
Short-range order is associated with exponential decay~\(e^{-r/L}\),
while quasi-long-range order is indicated by algebraic decay~\(r^{-n}\).
We extract the peaks from~\(g(r)\), and plot them with fits to both
exponential and algebraic decay functions in figure~\ref{fig:grfits}. At
densities of \(0.76\) and higher, exponential decay clearly
underestimates correlations at large~\(r/\ell\). At the two lowest
densities shown (\(\rho = 0.72, 0.73\)), the power law overestimates the
correlations, which have a pronounced downward curvature over all~\(r\).
It is difficult to unambiguously identify a transition density; however,
from examination of quality of fits, our best estimate for the
transition from short-range to quasi-long-range translational order
is~\(\rho \approx 0.76\).

A subtler form of spatial ordering can be studied by measuring the
angles of the ``bonds'' connecting the centers of pairs of neighboring
particles. (The bond angle~\(\theta_{jk}\) does not depend on the
orientation of individual particles, as illustrated in the inset diagram
in figure~\ref{fig:bond_op},~left). The global bond orientation order
parameter measures \(m\)-fold lattice orientation symmetry, and is given
by: \[\Psi_m = \abs{\ang{e^{im\theta_{jk}}}_{jk}}\] where
\(\theta_{jk}\) is the bond angle of neighboring particles \(j\)
and~\(k\). The average is calculated over all neighboring pairs; for
\(m=4\), neighbors are the four nearest particles, and for \(m=6\),
neighbors are given by the Delaunay triangulation. \(\Psi_m\) are shown
in figure~\ref{fig:bond_op}~(left). We find that \(\Psi_6\) remains near
zero for all densities, indicating the absence of hexagonal order; thus
the hexagonal rotator phase~{[}19{]} is not found for hard squares. On
the other hand, \(\Psi_4\) shows a marked increase in square symmetry as
density increases.

The preceding observations of the translational degrees of freedom
indicate the onset of translational order near \(\rho \gtrsim 0.76\),
which follows four-fold bond-orientational order at lower densities. In
the next section, we report molecular orientational order, followed by a
discussion of spatial correlations of both positional and orientational
order.

\subsection{Particle Orientational
Order}\label{particle-orientational-order}

\begin{figure} \center
\includegraphics[width=\textwidth]{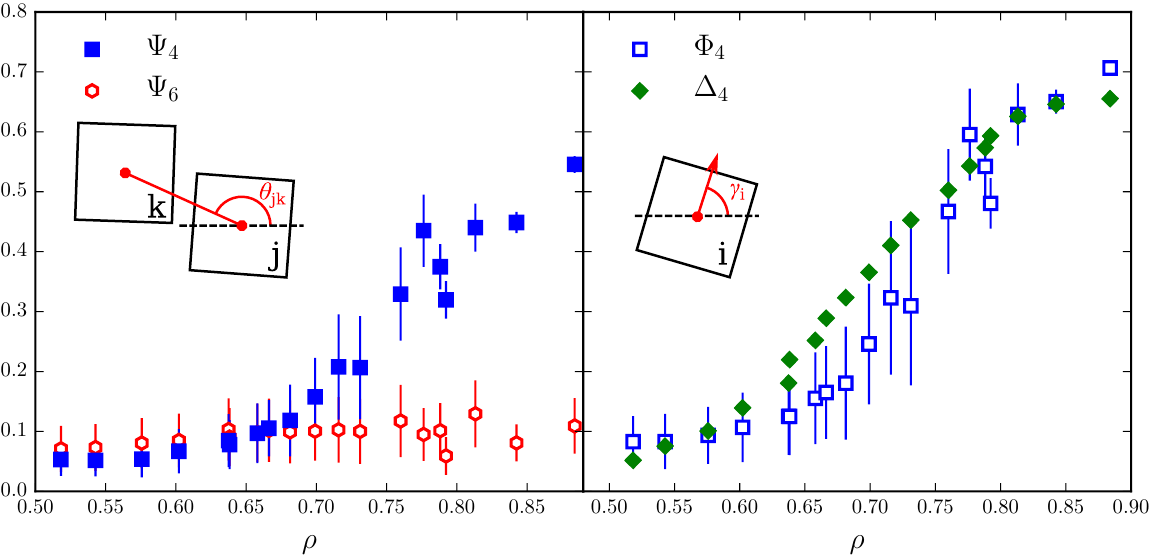}
\caption{Order Parameters. Left: The bond orientation order parameter $\Psi_m$ for both $m=4$ and $m=6$. Inset diagram shows the bond angle $\theta$. We see a clear increase in four-fold bond angle order with none for six-fold order, indicating an approach toward square crystalline order with no evidence for the hexagonal rotator phase at any density. Right: The particle orientation order parameters $\Phi_4$ and $\Delta_4$. Inset diagram shows the particle orientation angle $\gamma$. Both parameters show a parallel increase in four-fold particle orientation (tetratic) order, indicating a phase with stronger orientational order than translational order.}
\label{fig:op}
\label{fig:bond_op}
\label{fig:orient_op}
\end{figure}

\begin{figure} \center
\includegraphics[width=0.8\textwidth]{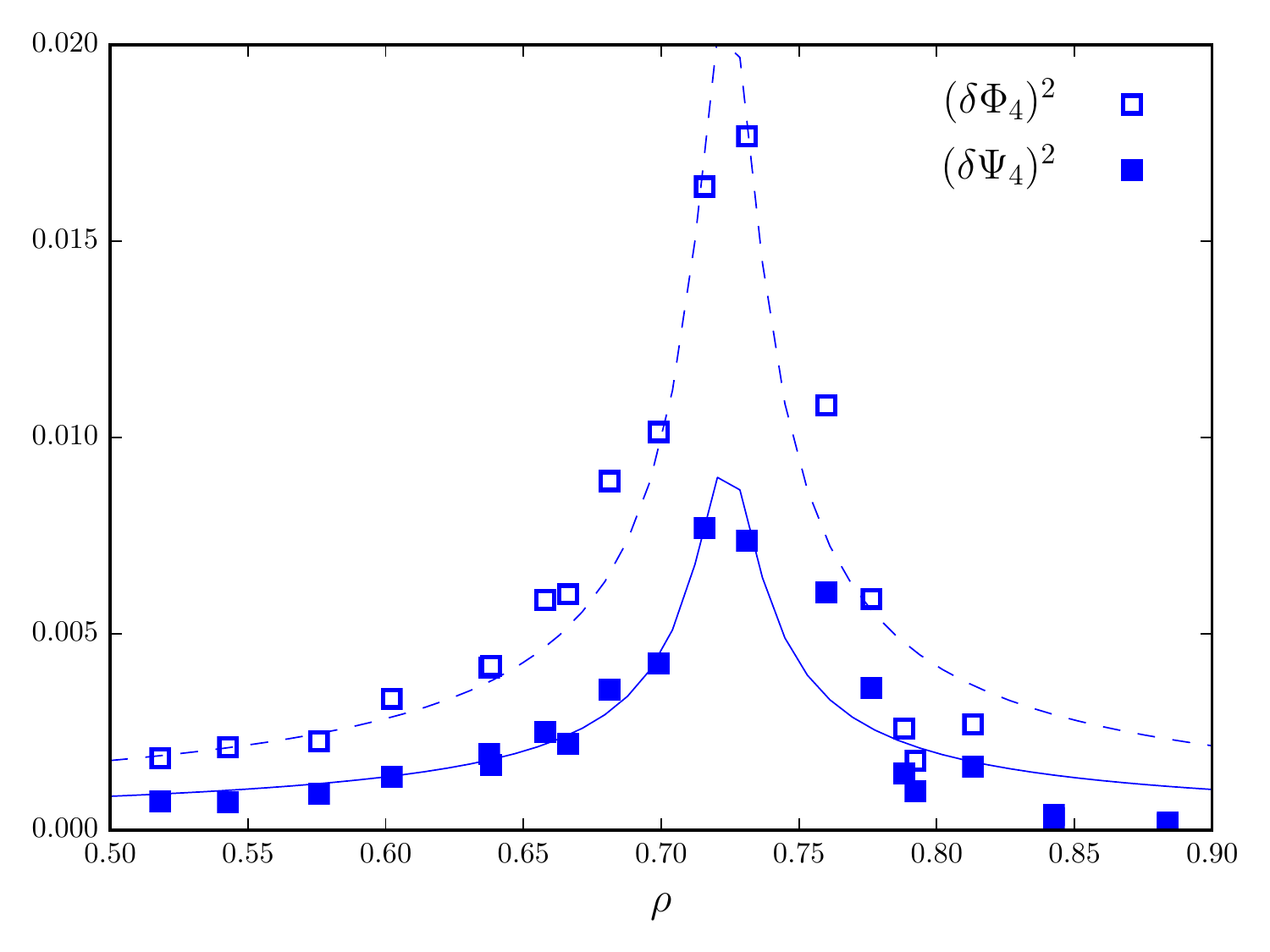}
\caption{Order Parameter Variances. The variances over long times of the order parameters shown in figure\ \ref{fig:op}. The peak in the variance indicates the existence of a phase transition. This occurs simultaneously at density $\rho_c = 0.72$ for both orientational and translational order, but the peak for orientational is stronger. The curves are best fits to $|\rho - \rho_c|^{-\eta}$, from which we extract the transition density. This indicates a coupling during this transition of these two degrees of freedom, dominated by orientational order.}
\label{fig:op_var}
\end{figure}

The ``molecular'' orientation of the individual particles is given by
the angle \(\gamma_i\) of the tile, illustrated in the inset diagram in
figure~\ref{fig:orient_op}~(right). Global orientational order of the
system is given by the particle orientation order parameter:
\[\Phi_m = \abs{\ang{e^{im\gamma_i}}_i}\] where \(m=4\) is the
rotational symmetry and the average is taken over all particles~\(i\).

We find that \(\Phi_4\) begins increasing above the noise at densities
in the range 0.64 to 0.68 with a marked rise above \(\rho \gtrsim 0.72\)
(figure~\ref{fig:orient_op},~right). This indicates an increase in
global tetratic (four-fold orientational) order, analogous to nematic
(two-fold) order.

A different, local measure of orientational alignment is the
distribution of orientation differences between neighboring particles,
as defined through a Delaunay triangulation. The width of this
distribution is the normalized standard deviation of these angle
differences: \[\Delta_m =
    1 - \frac{\sqrt{12}}{\pi/2}
        \sqrt{\ang{\left(\delta\gamma_{ij}\right)^2}_{\{i,j\}}}\] where
\(\delta\gamma_{ij}\) is the smallest angle between the orientations
(measured modulo \(2\pi/m\)) of neighboring particles \(i\) and \(j\),
and the average is over all such pairs. The values of \(\Delta_4\) agree
with the global orientational parameter \(\Phi_4\) as shown in
figure~\ref{fig:orient_op}, reflecting the concurrent emergence of local
and global orientational alignment.

Bond-orientation order \(\Psi_4\) appears at nearly the same density as
\(\Phi_4\). To better pin down the transition densities, we consider the
order parameter variances, which should diverge at the
transition~(figure~\ref{fig:op_var}). For both the particle- and
bond-orientation, the ensemble variance in the order parameter peaks
near density \(\rho = 0.72\), indicating a coupling between the
translational and orientational degrees of freedom. The smaller
amplitude of the bond-orientational order parameter suggests that the
transition may be driven by the ordering of particle orientations.

\begin{figure} \center
\includegraphics[width=\textwidth]{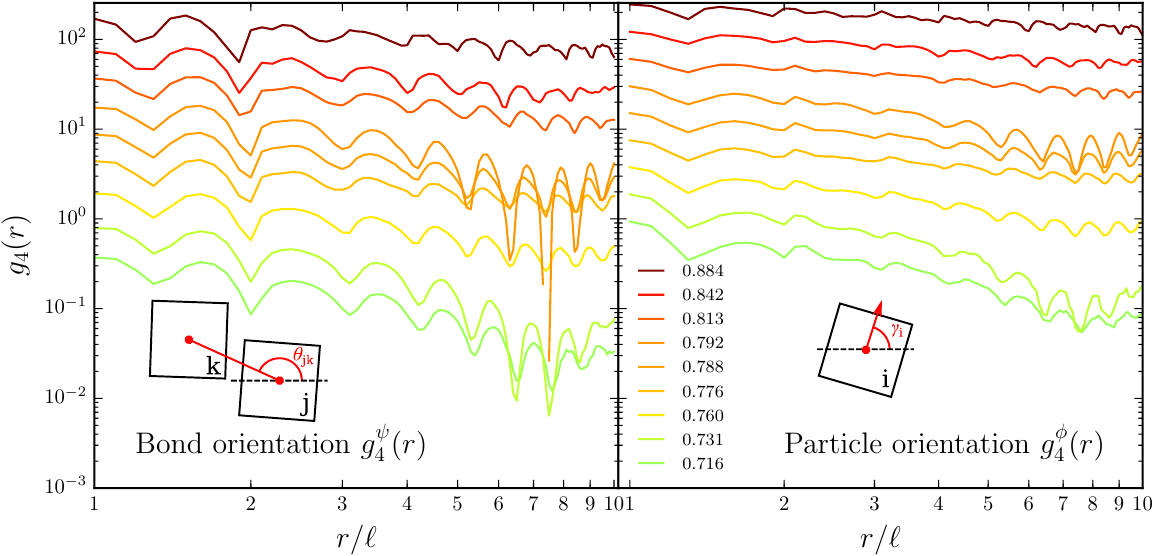}
\caption{Correlation functions of the order parameters. The correlation functions for bond orientation $g_m^\psi(r)$ (left) and particle orientation $g_m^\phi(r)$ (right). Curves of different density are vertically offset by a factor of 2 for clarity; the lowest curve has no offset. We find that bond orientation correlation has a lower amplitude and decays more quickly than the particle orientation correlation.}
\label{fig:corr}
\end{figure}

We measure the spatial extent of the above types of order by computing
the associated correlation functions. The form of the decay of each
correlation function allows us to distinguish between short-range order
and quasi-long-range order; the former is characterized by exponential
decay and the latter by algebraic decay.

The extent of particle orientational order is described by:
\[ g^\phi_m(r) = \ang{ \cos m \left(\gamma_i - \gamma_j\right)}_{r_{ij} = r} \]
where the average is taken over all pairs separated by distance~\(r\).
Our data is shown in figure~\ref{fig:corr},~left. We find that
exponential fits underestimate the correlation for densities above
\(\rho \approx 0.73\).

We compare this to the correlation function for the bond orientational
order parameter~\(\psi_m\), given by:
\[g^\psi_m(r) = \ang{\psi_m^*(r_j)\psi_m(r_k) }_{r_{ij}=r}\] Here
\(\psi_m(r_i)\) is the bond-orientational order parameter calculated for
the neighbors of particle \(i\). The average is taken over all pairs
separated by distance~\(r\). The form of this correlation
(figure~\ref{fig:corr},~right) matches that of \(g^\phi(r)\), but has
lower amplitude, consistent with the respective order parameters. We
observe significantly smaller oscillations in \(g^\phi(r)\) compared
with \(g^\psi(r)\). This indicates that bond-orientational order is weak
at positions incompatible with square-crystalline order, i.e., at
valleys in \(g(r)\), while molecular orientational order is less
sensitive to position.

\section{Dynamics}\label{dynamics}

\begin{figure} \center
\includegraphics[width=0.8\textwidth]{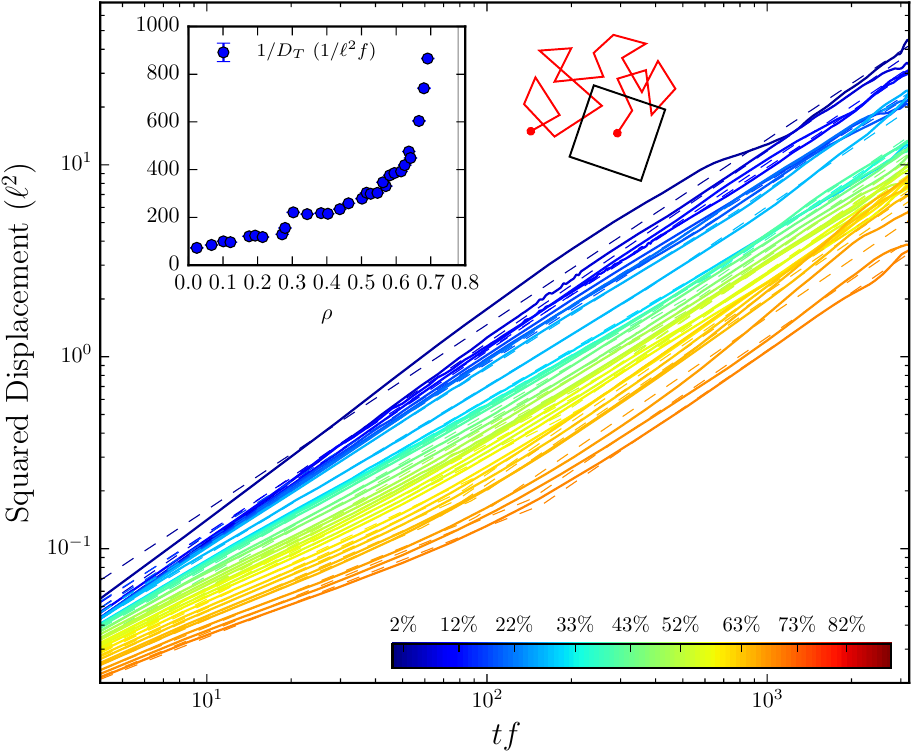}
\caption{Mean Squared Displacement (in units of particle size $\ell$) versus time (in units of vibration periods $1/f$). Particle motion at densities ranging from 0.03 to 0.70 shows diffusive behavior at low densities with a decreasing coefficient of diffusion. At densities above 0.30, motion is not diffusive at short times, becoming diffusive only after approximately $tf \approx 10^2$ vibration periods. Inverse Coefficient of Diffusion (inset): Fits for coefficient of diffusion $D_T$ are made to data for $tf > 500$.}
\label{fig:msd}
\end{figure}

\begin{figure} \center
\includegraphics[width=0.8\textwidth]{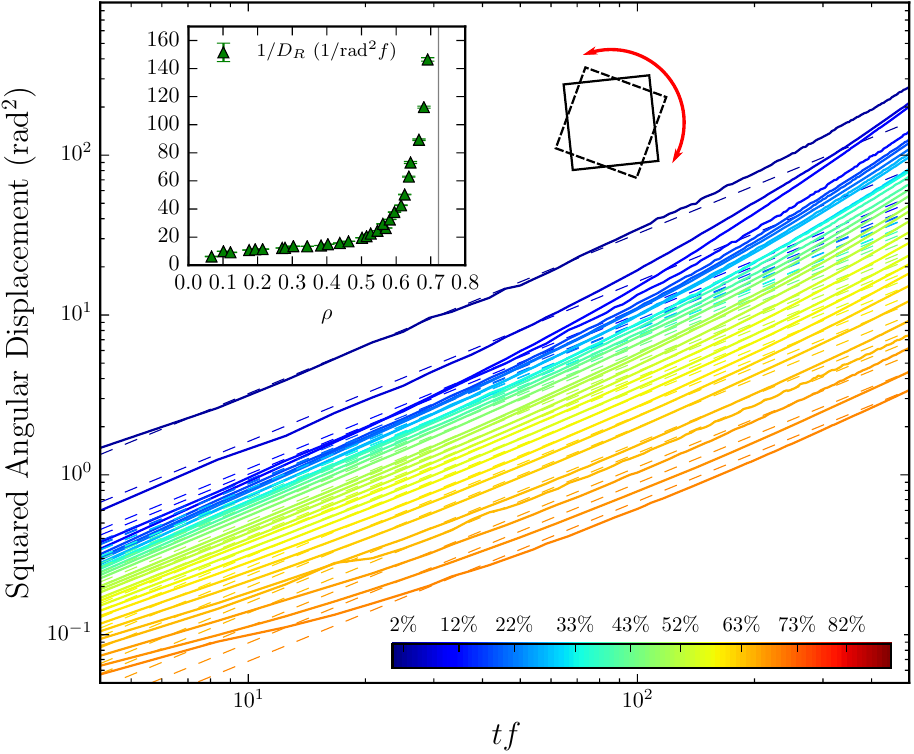}
\caption{Mean Squared Angular Displacement, with time shown in units of vibration periods $1/f$. Particle motion at densities ranging from 0.03 to 0.70 shows diffusive behavior at low densities with a decreasing coefficient of diffusion. The mean squared angular displacement is diffusive at short times, but at long times rotational motion is super-diffusive. Inverse Coefficient of Angular Diffusion (inset): Fits for coefficient of angular diffusion $D_R$ are made to data for $tf < 100$.}
\label{fig:msad}
\end{figure}

\begin{figure} \center
\includegraphics[width=\textwidth]{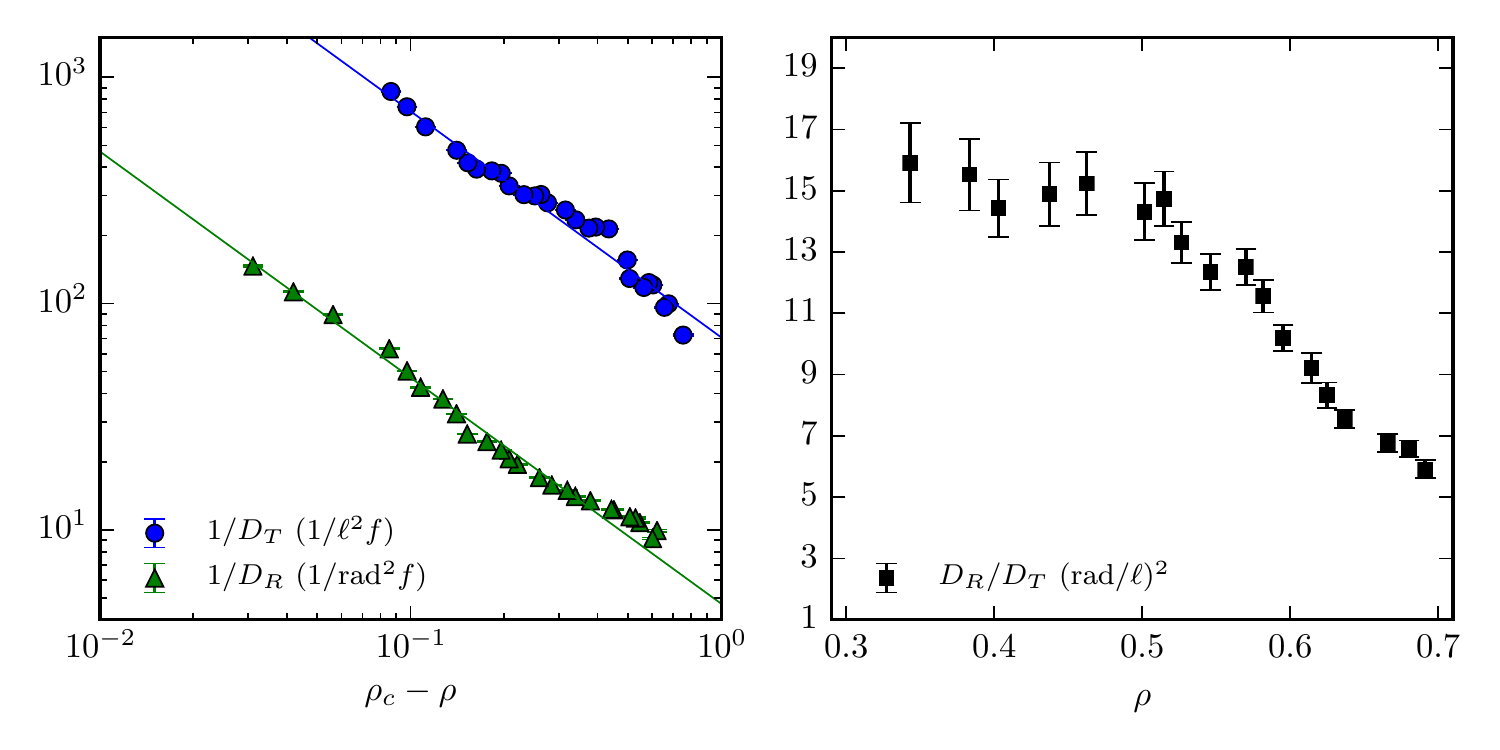}
\caption{Coefficients of Translational and Rotational Diffusion. Left: The coefficients plotted on a log-log scale versus $\rho_c - \rho$, showing power-law divergence. The values $\rho_c^R = 0.72$ and $\rho_c^T = 0.78$ are determined by best fit to $|\rho - \rho_c|^{-1}$. Right: the ratio of rotational to translational diffusion $D_R/D_T$ falls from its dilute limit as the density is increased, indicating a relative freezing of rotation compared with translation.}
\label{fig:dconst}
\label{fig:dratio}
\end{figure}

The rotational and translational dynamics of particles reflect the
constraints imposed by the spatial order in a phase. We track the
locations \(\vec x(t)\) and orientations \(\theta(t)\) of particles as a
function of time, from which we compute the mean squared displacement:
\[\Delta\vec x^2(t) = \ang{ \left[ \vec x_i(t_0 + t) - \vec x_i(t_0) \right]^2 }_{t_0,i}\]
where the average is taken over initial times \(t_0\) and all particles
\(i\). The mean squared angular displacement is defined similarly:
\[\Delta\theta^2(t) = \ang{ \left[ \theta_i(t_0 + t) - \theta_i(t_0) \right]^2 }_{t_0,i}\]
From each mean squared displacement, translational and rotational, we
obtain the coefficients of diffusion \(D_T\) and \(D_R\) by fitting a
power law of the form \(D t\) to the data within the diffusive regime
(figures~\ref{fig:msd} and~\ref{fig:msad}). As density is increased,
diffusion is slowed by collisions with neighboring particles. Both
coefficients decrease with increasing density, as shown in the plots of
\(1/D\) in the insets of figures~\ref{fig:msd} and \ref{fig:msad}. At
higher densities, particles become caged for short times and show a
sub-diffusive regime before the long-time behavior becomes
diffusive.\footnote{At very long times, we observe super-diffusive
  behavior, which results from a weak preference for one direction or
  the other in some particles. Rotation is unbiased when averaged over
  all particles.} As the density approaches a critical value, the
behavior of the diffusion coefficients is consistent with a power law,
shown plotted against \(\rho - \rho_c\) in figure~\ref{fig:dconst},
left.

To investigate the difference between translational and orientational
order, we compare \(D_T\) and \(D_R\). The rotational diffusion slows as
it approaches \(\rho_c^R = 0.72\), which is the same density at which
the orientational order parameter variance peaks
(figure~\ref{fig:op_var}). The translational diffusion freezes at
\(\rho_c^T = 0.78\), coinciding with the emergence of square crystalline
signatures in \(g(r)\) (figure~\ref{fig:gr}). Their relative evolution
can be seen directly without fitting by considering the decreasing ratio
\(D_R/D_T\) above \(\rho \approx 0.55\), shown in
figure~\ref{fig:dratio}, right. Thus the onset of order constrains
rotation at lower densities than it constrains translation.

\section{Conclusion}\label{conclusion}

In our system of vibrated hard squares, increasing density gives rise to
an increasing degree of order in the particle positions and
orientations. At low density, packings are positionally and
orientationally isotropic. Increasing density first leads to tetratic
orientational order in the range \(0.72 \lesssim \rho \lesssim 0.77\).
This phase is characterized by quasi-long-range, four-fold order in
particle orientation with short-range translational order. With further
increases in density, beyond \(\rho \approx 0.77\), the system tends
toward a quasi-long-range translationally ordered square phase. No
evidence is found for hexagonal positional order at any density.

\begin{table}[h]
\centering
\begin{tabular}{@{}ccclc@{}}
\toprule
Molec. & Bond & Transl. & Phase                                 & Density range  \\ \midrule
 ---   & ---  &   ---   & \multicolumn{1}{l|}{isotropic liquid} & 0.00--0.72     \\
  X    & ---  &   ---   & \multicolumn{1}{l|}{molecular orient} & ---            \\
 ---   &  X   &   ---   & \multicolumn{1}{l|}{bond-orient}      & ---            \\
  X    &  X   &   ---   & \multicolumn{1}{l|}{tetratic}         & 0.72--0.77     \\
 ---   &  X   &    X    & \multicolumn{1}{l|}{plastic (rotator) crystal} & ---   \\
  X    &  X   &    X    & \multicolumn{1}{l|}{square crystal}   & 0.77--1.00     \\
\bottomrule
\end{tabular}
\caption{Densities of observed phases. Phase boundary between tetratric and square crystal estimated at $\rho = 0.76$  and 0.78 by $g(r)$ decay and by translational diffusion $D_T$.}
\label{tab:dens}
\end{table}

Bond-orientational order, determined by particle positions alone,
increases in the same density range where molecular-orientational order
sets in, though with smaller amplitude and shorter-ranged correlations.
Though these are distinct kinds of order in independent degrees of
freedom, their coexistence is not surprising because position and
orientation are not separable in the pair interaction~{[}23,24{]}.

To our knowledge, this is the first experimental study of the
hard-square limit. This granular system displays a sequence of phases
consistent with equilibrium Monte Carlo simulations reported in~{[}12{]}
and~{[}18{]}. As in those studies, the limited system size restricts
structural analysis and smooths out transitions, making it difficult to
pinpoint transition densities or study critical behaviour. However,
measurements of the dynamics of the system --- inaccessible by Monte
Carlo --- strengthen the evidence for the sequence of phases. In
particular, the ratio of translational and rotational diffusion
constants serves as a useful tool to show that orientational freezing
precedes translational freezing.

\subsection{Acknowledgements}\label{acknowledgements}

We wish to thank JL Machta and JD Paulsen for helpful suggestions and
conversations. This work was supported by NSF-DMR 1207778 and 1506750.

\subsection{References}\label{references}

\hypertarget{refs}{}
\hypertarget{ref-KT}{}
{[}1{]} Kosterlitz J M and Thouless D J, 1973 \emph{J. Phys. C: Solid
State Phys.} \textbf{6} 1181

\hypertarget{ref-HN}{}
{[}2{]} Nelson D and Halperin B, 1979 \emph{Phys. Rev. B} \textbf{19}
2457--84

\hypertarget{ref-Y}{}
{[}3{]} Young A, 1979 \emph{Phys. Rev. B} \textbf{19} 1855--66

\hypertarget{ref-donev}{}
{[}4{]} Donev A, Burton J, Stillinger F and Torquato S, 2006 \emph{Phys.
Rev. B} \textbf{73} 054109

\hypertarget{ref-zhao-chaikin}{}
{[}5{]} Zhao K, Harrison C, Huse D, Russel W B and Chaikin P M, 2007
\emph{Phys. Rev. E} \textbf{76} 040401

\hypertarget{ref-cuesta}{}
{[}6{]} Cuesta J A and Frenkel D, 1990 \emph{Phys. Rev. A} \textbf{42}
2126--36

\hypertarget{ref-bates}{}
{[}7{]} Bates M and Frenkel D, 2000 \emph{J. Chem. Phys.} \textbf{112}
10034--41

\hypertarget{ref-narayan06}{}
{[}8{]} Narayan V, Menon N and Ramaswamy S, 2006 \emph{J Stat Mech}
\textbf{2006} P01005

\hypertarget{ref-muller}{}
{[}9{]} Müller T, de las Heras D, Rehberg I and Huang K, 2015
\emph{Phys. Rev. E} \textbf{91} 062207

\hypertarget{ref-geng-selinger}{}
{[}10{]} Geng J and Selinger J V, 2009 \emph{Phys. Rev. E} \textbf{80}
011707

\hypertarget{ref-mr-tetr}{}
{[}11{]} Martínez-Ratón Y and Velasco E, 2009 \emph{Phys. Rev. E}
\textbf{79} 011711

\hypertarget{ref-av-esc}{}
{[}12{]} Avendaño C and Escobedo F A, 2012 \emph{Soft Matter} \textbf{8}
4675--81

\hypertarget{ref-mr-geom}{}
{[}13{]} Martínez-Ratón Y, Velasco E and Mederos L, 2005 \emph{J. Chem.
Phys.} \textbf{122}

\hypertarget{ref-jiao-superdisks}{}
{[}14{]} Jiao Y, Stillinger F H and Torquato S, 2008 \emph{Phys. Rev.
Lett.} \textbf{100} 245504

\hypertarget{ref-rossi15}{}
{[}15{]} Rossi L, Soni V, Ashton D J, Pine D J, Philipse A P, Chaikin P
M, Dijkstra M, Sacanna S and Irvine W T M, 2015 \emph{PNAS} \textbf{112}
5286--90

\hypertarget{ref-batten}{}
{[}16{]} Batten R D, Stillinger F H and Torquato S, 2010 \emph{Phys.
Rev. E} \textbf{81} 061105

\hypertarget{ref-damasceno}{}
{[}17{]} Damasceno P F, Engel M and Glotzer S C, 2012 \emph{ACS Nano}
\textbf{6} 609--14

\hypertarget{ref-woj-frenk}{}
{[}18{]} Wojciechowski K W and Frenkel D, 2004 \emph{Comput. Methods}
\textbf{10} 235--55

\hypertarget{ref-zbm}{}
{[}19{]} Zhao K, Bruinsma R and Mason T G, 2011 \emph{PNAS} \textbf{108}
2684--7

\hypertarget{ref-reis}{}
{[}20{]} Reis P, Ingale R and Shattuck M, 2006 \emph{Phys. Rev. Lett.}
\textbf{96} 258001

\hypertarget{ref-galanis}{}
{[}21{]} Galanis J, Harries D, Sackett D, Losert W and Nossal R, 2006
\emph{Phys. Rev. Lett.} \textbf{96} 028002

\hypertarget{ref-safford}{}
{[}22{]} Safford K, Kantor Y, Kardar M and Kudrolli A, 2009 \emph{Phys.
Rev. E} \textbf{79} 061304

\hypertarget{ref-nelson-halperin-tilt}{}
{[}23{]} Nelson D and Halperin B, 1980 \emph{Phys. Rev. B} \textbf{21}
5312--29

\hypertarget{ref-GHB-MC}{}
{[}24{]} Gingras M, Holdsworth P and Bergersen B, 1990 \emph{Phys. Rev.
A} \textbf{41} 6786

\end{document}